\documentclass[final,5p,times,twocolumn]{elsarticle}
\usepackage{graphicx}
\usepackage{bm}
\usepackage{amsmath}
\usepackage{amsfonts}
\usepackage{amssymb}
\biboptions{compress}

\journal{Physics Letters B}

\begin{document}

\begin{frontmatter}

\title{Extraction of gluon distributions from structure functions at small $x$ in holographic QCD}

\author[IHEP,UCAS]{Akira Watanabe}

\ead{akira@ihep.ac.cn}

\author[OCU]{Takahiro Sawada}

\ead{tsawada@sci.osaka-cu.ac.jp}

\author[UCAS]{Mei Huang}

\ead{huangmei@ucas.ac.cn}

\address[IHEP]{Institute of High Energy Physics and Theoretical Physics Center for Science Facilities, Chinese Academy of Sciences, Beijing 100049, People's Republic of China}

\address[UCAS]{School of Nuclear Science and Technology, University of Chinese Academy of Sciences, Beijing 100049, People's Republic of China}

\address[OCU]{Department of Physics, Osaka City University, Osaka 558-8585, Japan}

\begin{abstract}
We investigate the nucleon and pion gluon distribution functions in the framework of holographic QCD, focusing on the small Bjorken $x$ region.
Based on an approximate relation, the gluon distributions are extracted from structure functions of the unpolarized deep inelastic scattering which can be calculated with a holographic QCD model, assuming the Pomeron exchange.
All the adjustable parameters of the model are determined with the HERA data of the proton structure functions.
We explicitly show that the extracted proton gluon distribution is consistent with results of the recent global QCD analysis.
The structure functions of the pion can be computed without any additional parameter, which enables us to predict its gluon distribution also.
We find that the resulting pion gluon density is smaller than the proton's, and agrees with the recent global QCD analysis result within the uncertainties.
\end{abstract}


\end{frontmatter}

\section{\label{sec:level1}Introduction}
Understanding the nucleon structure is one of the most important research topics in fundamental science, and tremendous efforts have been done to deepen our knowledge over several decades.
The quark-gluon structure of the nucleon is usually encoded into the parton distribution functions (PDFs) which are expressed with two kinematic variables, the Bjorken scaling variable $x$ and the probe scale $Q^2$.
The PDFs are required as inputs to make theoretical predictions or perform numerical simulations for high energy scattering phenomena.
The study of PDFs in the wide $x$ and $Q^2$ regime has provided us with great opportunities to improve our understandings about the various aspects of QCD.

To determine the PDFs, the global QCD analysis, utilizing high energy scattering data, is the most straightforward and reasonable way.
In particular, high statistics and high precision date of the deep inelastic scattering (DIS) have played a crucial role in such analysis.
Via the unpolarized lepton-nucleon DIS, the two independent structure functions, $F_2 (x, Q^2)$ and $F_L (x, Q^2)$, can be measured, and from these one can extract the information for PDFs, utilizing the formulae derived based on the perturbative technique of QCD.
Taking into account the recent high energy scattering data, including the DIS at HERA and the relevant events at LHC, various collaborations have performed the global QCD analysis to determine the proton PDFs~\cite{Harland-Lang:2014zoa,Abramowicz:2015mha,Alekhin:2017kpj,Ball:2017nwa,Hou:2019qau}.

Those efforts have gradually improved our knowledge about PDFs.
However, pinning down the gluon distribution, which is dominant in the small $x$ region, is still extremely difficult due to the nonperturbative nature of QCD.
To extract the gluon distribution, the precise measurement of the longitudinal structure function $F_L$ is essential.
In leading twist and next to leading order (NLO) QCD, $F_L$ can be expressed by the sum of two terms, which include $F_2$ and the gluon distribution function, respectively, and the contribution from the latter term is dominant at small $x$~\cite{Altarelli:1978tq}.
Since both $F_L$ and the gluon distribution function are highly nonperturbative physical quantities, in principle they are not calculable by the direct use of perturbative QCD.
Furthermore, although there are available $F_L$ data, those have large errors.
These facts cause the huge uncertainties which can be seen in the preceding studies based on the global QCD analysis.

In this work, we investigate the gluon distribution at small $x$ by calculating the DIS structure functions in the framework of holographic QCD~\cite{Kruczenski:2003be,Son:2003et,Kruczenski:2003uq,Sakai:2004cn,Erlich:2005qh,Sakai:2005yt,DaRold:2005zs,Brodsky:2014yha}, which is constructed based on the AdS/CFT correspondence~\cite{Maldacena:1997re,Gubser:1998bc,Witten:1998qj}, and applying an approximate relation between those structure functions and the gluon distribution.
It has been known that the Pomeron exchange picture well describes cross sections of various high energy scattering processes including DIS at small $x$~\cite{Donnachie:1992ny,ForshawRoss,PomeronPhysicsandQCD}.
The Pomeron exchange in QCD is interpreted as the Reggeized graviton exchange in the corresponding gravity theory.
Polchinski and Strassler first studied the high energy scattering based on the AdS/CFT correspondence~\cite{Polchinski:2001tt}, and subsequently a lot of related studies have been done~\cite{Polchinski:2002jw,BoschiFilho:2005yh,Brower:2006ea,Hatta:2007he,Brower:2007qh,BallonBayona:2007rs,Brower:2007xg,Cornalba:2008sp,Pire:2008zf,Cornalba:2010vk,Brower:2010wf,Watanabe:2012uc,Stoffers:2012zw,Watanabe:2013spa,Agozzino:2013zgy,Watanabe:2015mia,Watanabe:2018owy,Xie:2019soz}.
In particular, Brower, Polchinski, Strassler, and Tan (BPST) performed the gauge/string duality based analysis, and proposed a kernel which gives the Pomeron exchange contribution to high energy scattering cross sections~\cite{Brower:2006ea}.
This kernel has successfully been applied to the analysis of DIS at small $x$~\cite{Brower:2010wf,Watanabe:2012uc,Watanabe:2013spa,Agozzino:2013zgy,Watanabe:2015mia} and high energy hadron-hadron scattering~\cite{Watanabe:2018owy} so far.

In our model setup, the Pomeron exchange is described by the BPST kernel, and the wave functions of the U(1) vector field, which were derived in Ref.~\cite{Polchinski:2002jw}, is applied to describe the virtual photon in the five-dimensional AdS space.
For the Pomeron-nucleon coupling, we adopt the nucleon gravitational form factor which can be obtained from the bottom-up AdS/QCD model~\cite{Abidin:2008hn,Abidin:2009hr}.
There are four adjustable parameters in the model, which are to be determined with the HERA data for both the $F_2$ and $F_L$ structure functions, focusing on the highly nonperturbative kinematic regime, where $x \leq 10^{-2}$ and $Q^2 \leq 10$~GeV$^2$.
We will explicitly show that the considered data are well reproduced within the model.
Then, utilizing the approximate relation proposed by the authors of Ref.~\cite{CooperSarkar:1987ds}, we extract the gluon distribution from the resulting structure functions.
This extraction can be done without any additional parameter.
It will be presented that our prediction is consistent with results of the recent global QCD analysis.

Furthermore, we also investigate the gluon distribution of the pion in the framework.
Once all the model parameters are determined by the fit with the proton DIS data, the pion structure functions can be calculated without any additional parameter in the chiral limit.
Hence, the pion gluon distribution can also be predicted, and we find that the resulting gluon density is smaller than the proton's.
Recently, the first Monte Carlo global analysis of the pion PDFs was performed~\cite{Barry:2018ort}.
It will be shown that our prediction agrees with their result within the uncertainties.

\section{\label{sec:level2}Theoretical framework}
In the quark-parton model, the longitudinal structure function $F_L (x, Q^2)$ vanishes, which means that the origin of the experimentally observed finite $F_L$ is higher order effects of QCD.
In leading twist and NLO QCD, $F_L$ is expressed in terms of $F_2$ and the gluon distribution function $g(x, Q^2)$ by~\cite{Altarelli:1978tq}
\begin{align}
&{F_L}\left( {x,{Q^2}} \right) = \frac{{{\alpha _s}\left( {{Q^2}} \right)}}{{4\pi }}\left( {\frac{{16}}{3}{I_F} + 8\sum\limits_i {e_i^2} {I_G}} \right) ,
\label{eq:AM_eq} \\
&{I_F} = \int_x^1 {\frac{{dy}}{y}} {\left( {\frac{x}{y}} \right)^2}{F_2}\left( {y,{Q^2}} \right),
\label{eq:IF_original} \\
&{I_G} = \int_x^1 {\frac{{dy}}{y}} {\left( {\frac{x}{y}} \right)^2}\left( {1 - \frac{x}{y}} \right) \mathcal{G} \left( {y,{Q^2}} \right),
\label{eq:IG_original}
\end{align}
where $\mathcal{G} (y, Q^2) = y g(y, Q^2)$ and $\alpha _s$ and $e_i$ are the QCD coupling and the quark charge, respectively.

Substituting $y = x / (1 - z)$, Eqs.~\eqref{eq:IF_original} and~\eqref{eq:IG_original} are rewritten as
\begin{align}
&{I_F} = \int_0^{1 - x} {dz} \left( {1 - z} \right){F_2}\left( {\frac{x}{{1 - z}}} \right),
\label{eq:IF_rewritten} \\
&{I_G} = \int_0^{1 - x} {dz} z\left( {1 - z} \right) \mathcal{G} \left( {\frac{x}{{1 - z}}} \right),
\label{eq:IG_rewritten}
\end{align}
respectively.
When $x$ is small enough, $F_2$ and $\mathcal{G}$ in the above equations can be expanded.
Hence, following the procedures presented in Ref.~\cite{CooperSarkar:1987ds}, $I_F$ and $I_G$ are expressed as $I_F \approx F_2 (2x) / 2$ and $I_G \approx \mathcal{G} \left( x / 0.4 \right) / 5.9$, respectively.
Therefore, one obtains the approximate relation between the gluon distribution function and the structure functions:
\begin{align}
&xg\left( {x,{Q^2}} \right) \nonumber \\
&\approx 5.9 \left[ \frac{\pi }{{2{\alpha _s}\left( {{Q^2}} \right)}}{F_L}\left( {0.4x,{Q^2}} \right) - \frac{1}{3}{F_2}\left( {0.8x,{Q^2}} \right) \right] / \sum\limits_i {e_i^2}.
\label{eq:xg_approximated}
\end{align}
Due to the relatively large magnitude of the coefficient multiplying $F_L$, it is understood that the gluon distribution is mainly dependent on $F_L$.
However, since the contribution from $F_2$ term is not negligibly small, we consider the both structure functions to numerically evaluate the gluon distribution.

In this study, we calculate those structure functions within a model, and extract the gluon distribution by applying Eq.~\eqref{eq:xg_approximated}.
We take into account the small $x$ region down to $x = 10^{-6}$ which is significantly smaller compared to the $x$ range that the authors of Ref.~\cite{CooperSarkar:1987ds} considered.
At such a small $x$, it is possible that the gluon saturation effect becomes visible, which may affect validity of the approximate relation.
However, from the currently available HERA data one has not explicitly seen the effect.
Hence, in the present work we apply the same approximate relation to extract the gluon distribution.
As to the $Q^2$ value, we choose a typical reference scale $Q^2 = 10$~GeV$^2$, and at this scale we compare the resulting gluon distributions with recent results of the global QCD analysis.

The DIS structure functions at small $x$ can be calculated in the framework of holographic QCD, assuming the Pomeron exchange in the five-dimensional AdS space.
Employing the BPST kernel denoted by $\chi$~\cite{Brower:2006ea}, the scattering amplitude of a two-body process, $1 + 2 \to 3 + 4$, is expressed in the eikonal representation as
\begin{align}
\mathcal{A} (s, t) = &2 i s \int d^2 b e^{i \bm{k_\perp } \cdot \bm{b}} \int dz dz' \nonumber \\
&\times P_{13} (z) P_{24} (z') \left[ 1-e^{i \chi (s, \bm{b}, z, z')} \right],
\label{eq:a}
\end{align}
where
$s$ and $t$ are the Mandelstam variables, %
$\bm{ b }$ is the two-dimensional impact parameter, %
$z (z')$ are the fifth coordinates for the incident(target) particles, %
and
$P_{13}(z)$ and $P_{24}(z')$ represent the density distributions of the involved two particles in the AdS space.

To obtain the structure functions, we consider the total cross section of the forward scattering by applying the optical theorem.
Keeping only the leading contribution from the kernel, the Pomeron exchange contribution is expressed by the imaginary part of the kernel.
In the conformal limit, the analytical form of $\mbox{Im} \chi$ can be obtained, and the impact parameter integration in Eq.~\eqref{eq:a} can also be performed analytically~\cite{Brower:2006ea,Brower:2007xg}.
Hence, the structure functions are written as
\begin{align}
&F_i (x, Q^2) = \frac{g_0^2 \rho^{3/2} Q^2}{32 \pi ^{5/2}} %
\int dz dz' P_{13}^{(i)} (z, Q^2) P_{24} (z') \nonumber \\
&\hspace{41mm} \times (z z') \mbox{Im} [ \chi_{c} (s, z, z') ],
\label{eq:SF} \\
&\mbox{Im} [\chi_c(s, z, z') ] \equiv %
e^{(1-\rho) \tau} e^ {- \left[ ({\log ^2 z / z'})/{\rho \tau} \right]} / {\tau^{1/2}},
\label{eq:kernel_conformal}
\end{align}
where $i = 2, L$ and $\tau = \log (\rho z z' s / 2)$.
$g_0^2$ and $\rho$ are adjustable parameters which control the magnitude and the energy dependence of the structure functions, respectively.

In the preceding studies~\cite{Brower:2010wf,Watanabe:2012uc,Watanabe:2013spa}, it was shown that the inclusion of the confinement effect in QCD is necessary to reproduce the proton structure function data with the BPST kernel, unless we consider the high $Q^2$ region.
Therefore, instead of the conformal kernel, to numerically evaluate the structure functions we employ the modified kernel:
\begin{align}
&\mbox{Im} [ \chi_{mod} (s, z, z') ] \equiv %
\mbox{Im} [ \chi_c (s, z, z') ] \nonumber \\
&\hspace{30mm} + \mathcal{F} ( s, z, z' ) %
\mbox{Im} [ \chi_c (s, z, z_0 z_0' / z') ],
\label{eq:kernel_modified} \\
&\mathcal{F} ( s, z, z' ) = 1 - 2 \sqrt{ \rho \pi \tau } %
e^{ \eta^2 } \mbox{erfc} ( \eta ),
\label{eq:mathcal_F} \\
&\eta = \left( - \log \frac{ z z' }{ z_0 z_0' } %
+ \rho \tau \right) / { \sqrt{ \rho \tau }},
\label{eq:eta}
\end{align}
where $z_0 (z'_0)$ are the cutoffs of the fifth coordinates.
Note that $z_0$ is one of the adjustable parameters of the model, but $z'_0$ is uniquely fixed with hadron masses.
The first term in the right-hand side of Eq.~\eqref{eq:kernel_modified} is exactly the same as the conformal kernel, and the second term mimics the confinement effect with the same functional form.

To numerically evaluate the structure functions, one needs to specify the density distributions, $P^{(i)}_{13}(z, Q^2)$ and $P_{24}(z')$ in Eq.~\eqref{eq:SF}.
For the probe photon, we apply the wave functions of the five-dimensional U(1) vector field with a weight $w$ on its longitudinal component:
\begin{align}
&P_{13}^{(2)}(z,Q^2) = Q^2 z \left[ w K_0^2(Qz) + K_1^2(Qz) \right],
\label{eq:P13_2} \\
&P_{13}^{(L)}(z,Q^2) = w Q^2 z K_0^2(Qz),
\label{eq:P13_L}
\end{align}
where $K_{0(1)}$ are the modified Bessel functions of the second kind.
When $w = 1$, the results presented in Ref.~\cite{Polchinski:2002jw}, which satisfy the Maxwell equation in the bulk AdS space, are recovered.
It has been known that their results are useful to reproduce the experimental data of proton $F_2$, however, the resulting $F_L$ is somewhat larger than the data.
This can be seen in the observations that the experimentally favored value of the longitudinal-to-transverse ratio, $R = F_L / F_T = 0.26$~\cite{Collaboration:2010ry}, is obviously smaller compared to the theoretical predictions, $0.3 \lesssim R \lesssim 0.5$~\cite{Cornalba:2010vk,Brower:2010wf,Watanabe:2013spa,Watanabe:2015mia}.
Hence, as a simple ansatz, we introduce the weight, which is to be determined with the data, in this study.

For the density distributions of the target hadrons, $P_{24}(z')$ in Eq.~\eqref{eq:SF}, we apply the gravitational form factors which can be obtained with the bottom-up AdS/QCD models~\cite{Abidin:2008hn,Abidin:2009hr}.
As discussed in Refs.~\cite{Henningson:1998cd,Muck:1998iz,Contino:2004vy,Hong:2006ta}, in the model the nucleon is described as a solution to the five-dimensional Dirac equation.
The density distribution of the proton is given in terms of the Bessel functions by~\cite{Abidin:2009hr}
\begin{align}
&P_p (z) = \frac{1}{2 z^{3}} %
\left[ \psi_L^2 (z) + \psi_R^2 (z) \right],
\label{eq:P24_proton} \\
&\psi_L (z) = \frac{\sqrt{2} z^2 J_2 (m_p z)}{z_0^p J_2 (m_p z_0^p)}, \ \
\psi_R (z) = \frac{\sqrt{2} z^2 J_1 (m_p z)}{z_0^p J_2 (m_p z_0^p)},
\label{eq:wavefunctions_proton}
\end{align}
where $\psi_{L(R)}$ are the left-handed and right-handed components of the Dirac field, respectively, and the cutoff parameter $z_0^p$ is fixed with the proton mass $m_p$ by the condition, $J_1 (m_p z_0^p) = 0$.
Utilizing the bottom-up AdS/QCD model of mesons~\cite{Erlich:2005qh}, the analytical form of the pion wave function can be obtained in the chiral limit.
The density distribution of the pion is expressed with Bessel functions as~\cite{Abidin:2008hn}
\begin{align}
&P_\pi ( z ) = \frac{ \left[ \partial _{z} \Psi ( z ) \right] ^2 }{4 \pi^2  f_\pi ^2 z} %
+ \frac{\sigma^2  z^6 \Psi (z)^2 }{ f_\pi ^2  z^3 },
\label{eq:P24_pion} \\
&\Psi \left(  z \right) = %
z \Gamma \left[  {\frac{2}{3}} \right] \left(  {\frac{\alpha }{2}} \right) ^{1/3} %
\Biggl[  I_{ - 1/3} \left(  {\alpha z^3 } \right) \nonumber \\
&\hspace{27mm} - I_{1/3} \left(  {\alpha z^3 } \right) %
\frac{{I_{2/3} \left( {\alpha (z_0^\pi ) ^3 } \right) }}{{I_{ - 2/3} \left(  {\alpha (z_0^\pi ) ^3 } \right) }}  \Biggr] ,
\label{eq:wavefunctions_pion}
\end{align}
where $f_\pi $ is the pion decay constant, %
$\alpha = 2 \pi \sigma / 3$, %
and
$\sigma = (332$~MeV$)^3$ is used in this study.
$z_0^\pi$ is the cutoff parameter which is uniquely fixed with the $\rho $ meson mass $m_\rho $ by the condition, $J_0 ( m_\rho z_0^\pi ) = 0$.

\section{\label{sec:level3}Numerical results}
In the model setup, there are four adjustable parameters in total, which are determined by a numerical fit, considering the HERA data for both the proton $F_2$~\cite{Aaron:2009aa} and $F_L$~\cite{Collaboration:2010ry} structure functions simultaneously in the kinematic regime, where $x \leq 10^{-2}$ and $Q^2 \leq 10$~GeV$^2$.
For this procedure, the MINUIT package~\cite{James:1975dr} is used.
The resulting best fit values of the parameters are found to be:
$g_0^2 = 162.15 \pm 6.71$, $\rho = 0.7798 \pm 0.0021$, $z_0 = 3.059 \pm 0.079$, $w = 0.6198 \pm 0.0923$, and the chi-square value per degree of freedom is $\chi_{d.o.f.}^2 =$ 1.295 (240.8/186).

\begin{figure}[bt]
\begin{center}
\includegraphics[width=0.43\textwidth]{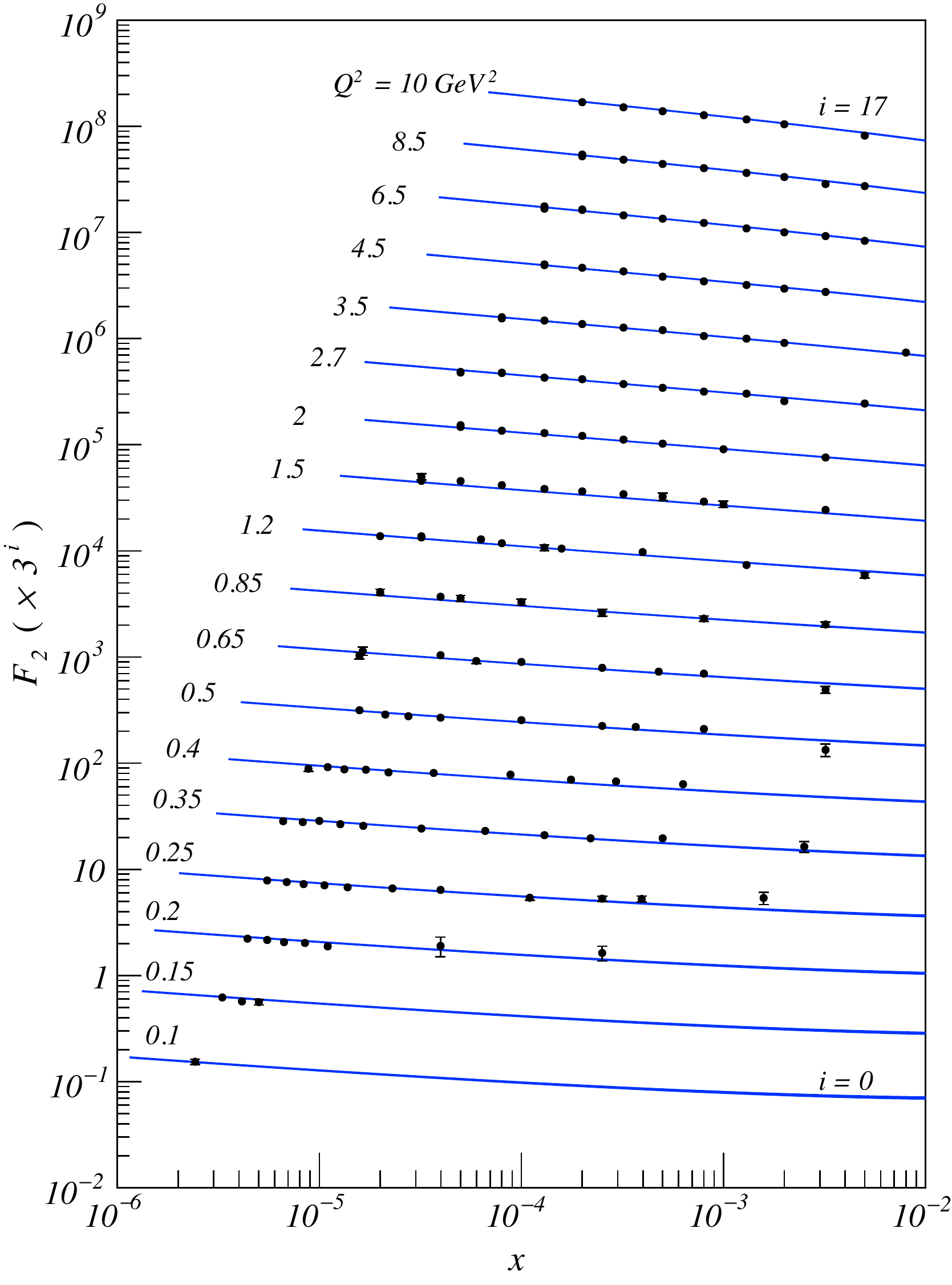}
\caption{
$F_2^p (x, Q^2)$ as a function of the Bjorken $x$ for various $Q^2$.
Our calculations are denoted by solid curves with blue bands representing the uncertainties.
The HERA data~\cite{Aaron:2009aa} are depicted by circles with error bars.
For a display purpose, the absolute magnitudes are scaled by a factor $3^i$.
}
\label{fig:F2_proton}
\end{center}
\end{figure}

\begin{figure}[bt]
\begin{center}
\includegraphics[width=0.47\textwidth]{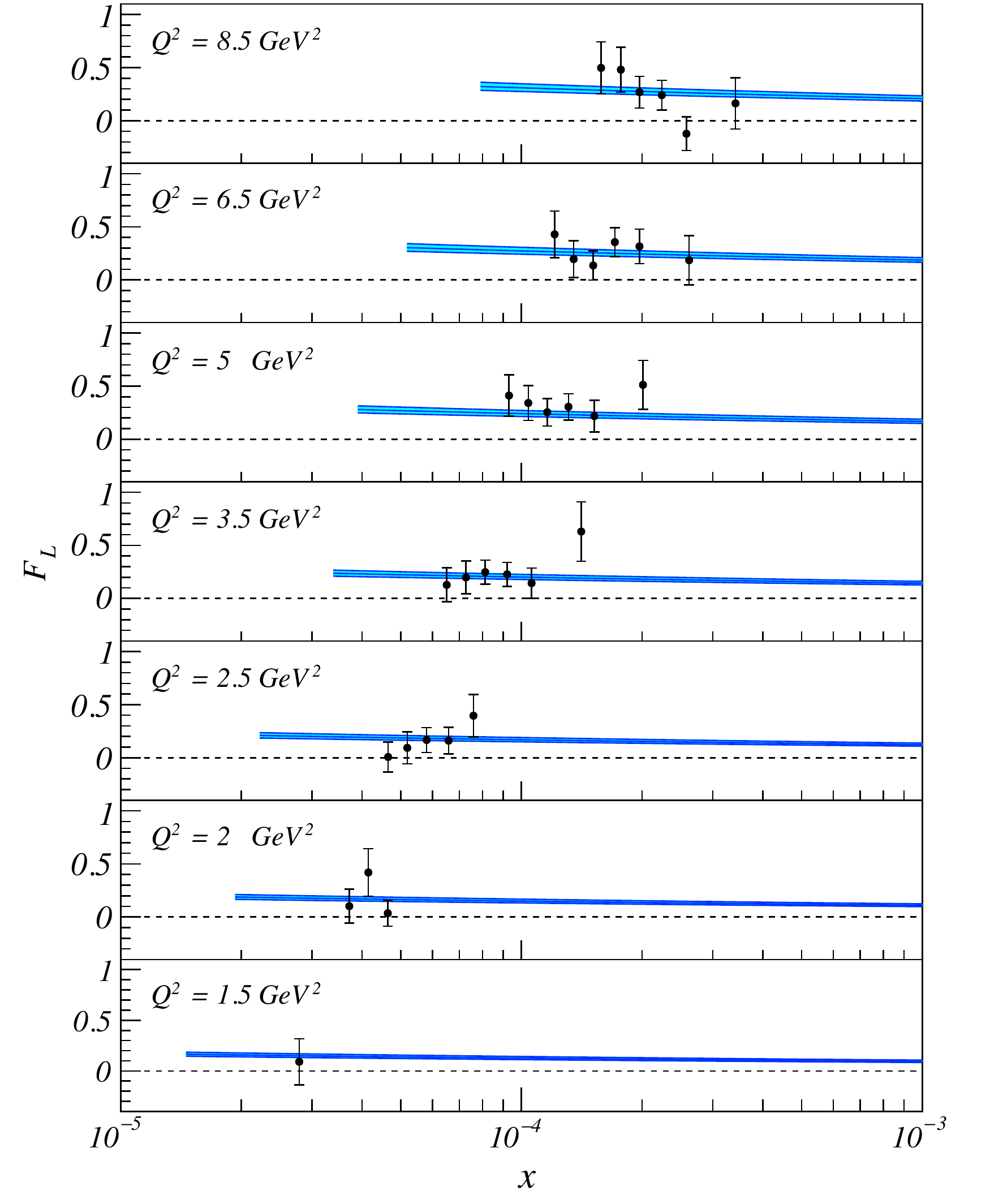}
\caption{
$F_L^p (x, Q^2)$ as a function of the Bjorken $x$ for various $Q^2$.
Our calculations are denoted by solid curves with blue bands representing the uncertainties.
The HERA data~\cite{Collaboration:2010ry} are depicted by circles with error bars.
}
\label{fig:FL_proton}
\end{center}
\end{figure}

We show in Fig.~\ref{fig:F2_proton}
and Fig.~\ref{fig:FL_proton}
the resulting $F_2$ and $F_L$ structure functions of the proton, respectively, compared to the HERA data.
It is seen that the both $F_2$ and $F_L$ data in the whole considered kinematic regime are well described by the model.

\begin{figure}[bt]
\begin{center}
\includegraphics[width=0.47\textwidth]{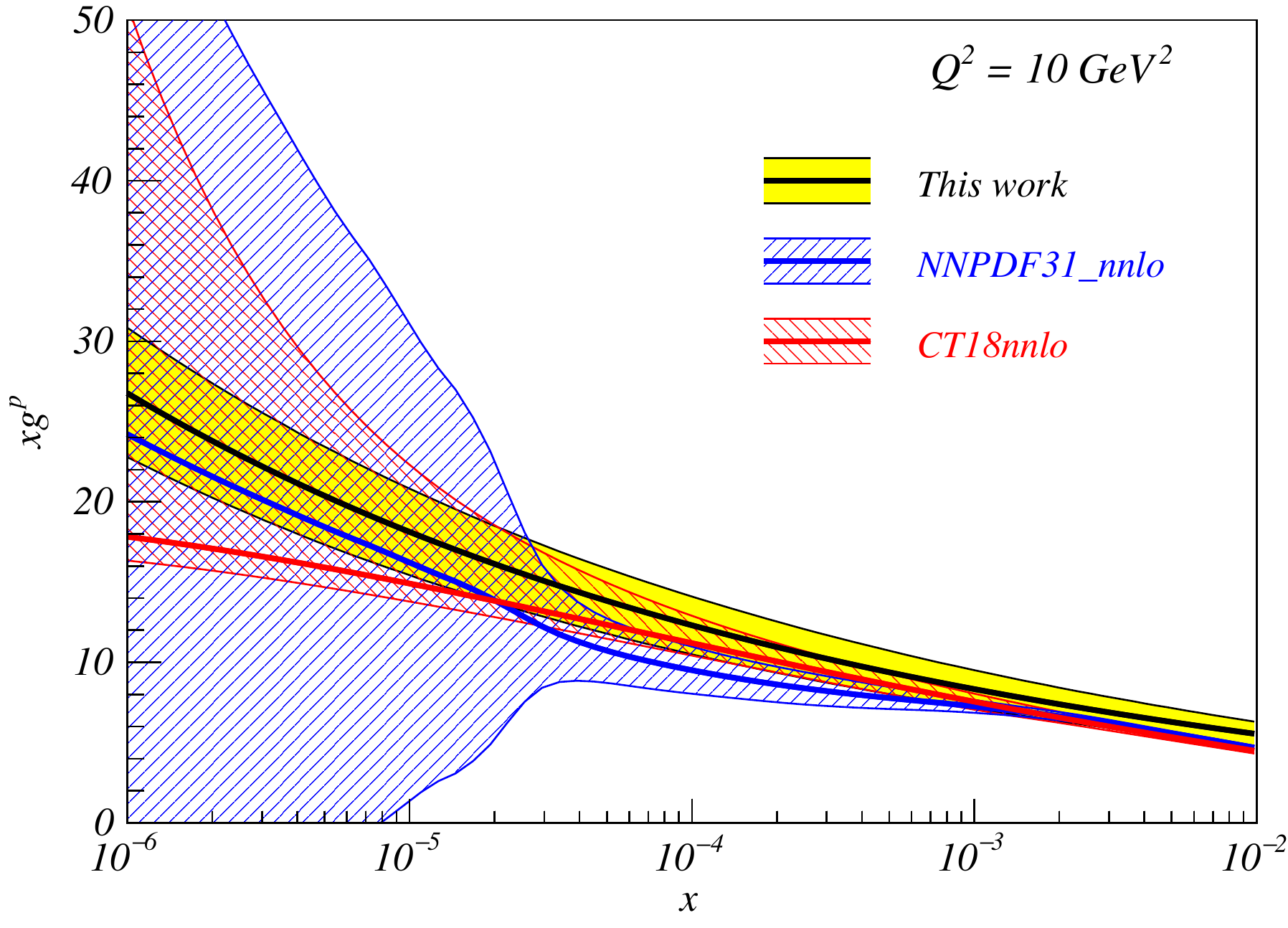}
\caption{
The proton gluon distribution $xg^p (x, Q^2)$ as a function of the Bjorken $x$ at $Q^2 = 10$~GeV$^2$.
The yellow, blue, and red bands represent our calculation and the global QCD analysis results of the NNPDF~\cite{Ball:2017nwa} and CTEQ-TEA~\cite{Hou:2019qau} collaborations, respectively, with the 68\% C.L. uncertainties.
}
\label{fig:xg_proton}
\end{center}
\end{figure}

Once all the four model parameters are determined with the data, we can extract the proton gluon distribution via Eq.~\eqref{eq:xg_approximated} from the obtained structure functions.
To do this, we consider the four flavor case, and utilize the approximate NLO solution to the renormalization group equation to calculate $\alpha _s (Q^2)$ (see Ref.~\cite{Furmanski:1981cw} for instance) in this study.
We display in Fig.~\ref{fig:xg_proton}
the $x$ dependence of the resulting proton gluon distribution at $Q^2 = 10$~GeV$^2$, compared to the recent global QCD analysis results obtained by the NNPDF~\cite{Ball:2017nwa} and CTEQ-TEA~\cite{Hou:2019qau} collaborations.
One can see from the figure that our calculation is consistent with those results in the whole considered region within the uncertainties.
It should be noted that the error band of our result purely reflects the errors of the HERA data for the proton $F_2$ and $F_L$, and the uncertainties originated from the approximate relation Eq.~\eqref{eq:xg_approximated} and the model itself are not included.
Further remarks on this will be given in the next section.

\begin{figure}[bt!]
\begin{center}
\includegraphics[width=0.47\textwidth]{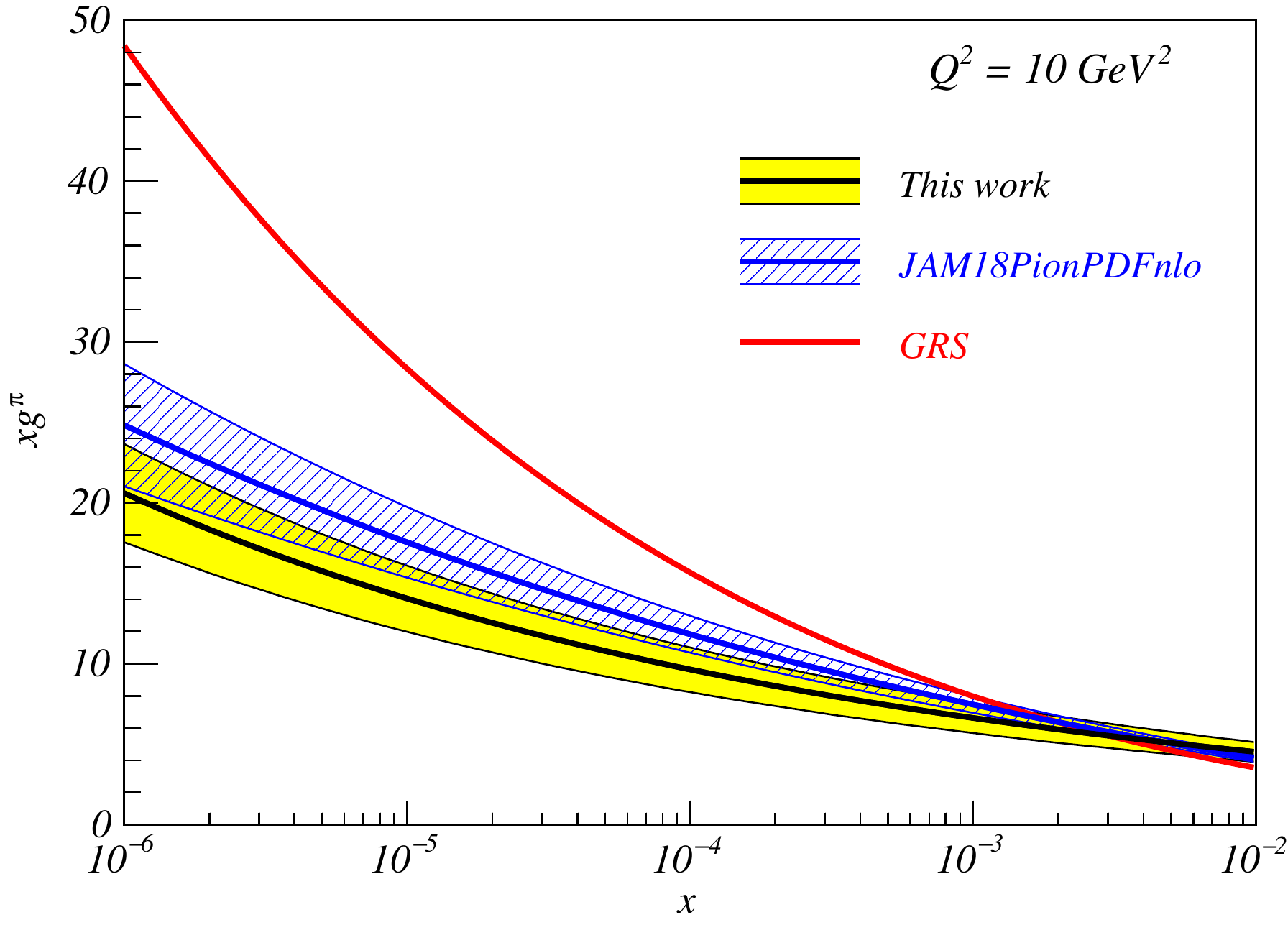}
\caption{
The pion gluon distribution $xg^\pi (x, Q^2)$ as a function of the Bjorken $x$ at $Q^2 = 10$~GeV$^2$.
The yellow and blue bands represent our calculation and the global QCD analysis result of the JAM collaboration~\cite{Barry:2018ort}, respectively, with the 68\% C.L. uncertainties.
The red curve depicts the result of GRS~\cite{Gluck:1999xe}.
}
\label{fig:xg_pion}
\end{center}
\end{figure}

Using Eq.~\eqref{eq:P24_pion}, instead of Eq.~\eqref{eq:P24_proton}, as the target hadron density distribution, we can obtain the pion structure functions, and then extract its gluon distribution $xg^\pi (x, Q^2)$ without any additional parameter.
We present our prediction at $Q^2 = 10$~GeV$^2$ in Fig.~\ref{fig:xg_pion}.
It is seen from the comparison that our calculation agrees with the global QCD analysis result obtained by the JAM collaboration~\cite{Barry:2018ort} in the whole considered $x$ region within the uncertainties, although the result obtained from the parameterization proposed by Gl\"uck, Reya, and Schienbein (GRS)~\cite{Gluck:1999xe} is substantially larger than the others.
Also, it is found that in our results the mean value of $xg^\pi$ is smaller than that of $xg^p$ in the considered $x$ range.

\section{\label{sec:level4}Summary and discussion}
We have investigated the nucleon and pion gluon distribution functions at small Bjorken $x$ in the framework of holographic QCD.
The gluon distributions were extracted via an approximate relation from the unpolarized DIS structure functions which were calculated within a holographic QCD model, assuming dominance of the Pomeron exchange in the five-dimensional AdS space.
All the model parameters were determined with the HERA data for the proton structure functions, and the pion structure functions and its gluon distribution were predicted without any additional parameter.

The resulting proton gluon distribution agrees with results of the recent global QCD analysis.
In particular, at $x < 10^{-5}$ the uncertainty of our result is significantly smaller than the other ones.
However, this does not imply that our approach can better constrain the distribution compared to the global QCD analysis, because in our approach there are systematic uncertainties which are not explicitly shown in the figure due to the difficulties in estimation.
Further investigations of validity of the approximate relation Eq.~\eqref{eq:xg_approximated} in the small $x$ region are needed to reduce the uncertainties.

It should be emphasized that our pion result mainly shows the advantage of our approach.
The pion gluon distribution has been poorly known in the literature, but meanwhile the JAM collaboration recently performed the first Monte Carlo global analysis of the pion PDFs.
Their gluon distribution was constrained mainly by the leading neutron electroproduction data, however, for this process there is no data at $x < 10^{-3}$.
In the present analysis, it was observed that the resulting pion gluon distribution is consistent with their result within the uncertainties.
Furthermore, within our model the resulting gluon density of the pion is smaller than that of the proton in the whole considered $x$ range.
This is a qualitative result, and may imply that there is a difference in the gluonic structure between the proton and the pion.
Further studies by different approaches are certainly required to pin down this.

Finally, we make remarks for possible improvements.
Although this is the first study of the gluon distributions in the framework of holographic QCD, we obtained reasonable results.
However, improving the model itself may be possible.
In this work, we only focused on the single-Pomeron exchange, and adopted the hard-wall AdS/QCD models to obtain the target hadron density distributions.
The multi-Pomeron exchange, the soft-wall versions, and the physical pion case shall be considered.
Also, selecting the considered $Q^2$ range may affect the results.
In fact, if we cut the low $Q^2$ ($\lesssim 1$~GeV$^2$) data and consider the slightly larger $Q^2$ ($\lesssim 20$~GeV$^2$) region, the resulting $\chi_{d.o.f.}^2$ is significantly improved.
Further studies would also help to better understand the applicable limit of the model.
Moreover, the importance of the further measurements of the longitudinal structure function should be emphasized.
Since the errors of $F_L$ data considered in this study are large, the lepton-nucleon DIS experiment is still necessary to deepen our understandings about the nucleon structure in the small Bjorken $x$ region.

\section*{Acknowledgements}
We thank Wen-Chen Chang, Robert Ciesielski, Tie-Jiun Hou, Hsiang-nan Li, and Katsuhiko Suzuki for useful discussions.
The work of A.W. was supported by Chinese Academy of Sciences President's International Fellowship Initiative under Grant No. 2019PM0124 and partially by China Postdoctoral Science Foundation under Grant No. 2018M641473.
The work of M.H. was supported in part by the NSFC under Grant Nos. 11725523, 11735007, 11261130311 (CRC 110 by DFG and NSFC), Chinese Academy of Sciences under Grant No. XDPB09, and the start-up funding from University of Chinese Academy of Sciences.

\bibliographystyle{elsarticle-num}

\bibliography{hQCD}

\begin{thebibliography}{10}
\expandafter\ifx\csname url\endcsname\relax
  \def\url#1{\texttt{#1}}\fi
\expandafter\ifx\csname urlprefix\endcsname\relax\def\urlprefix{URL }\fi
\expandafter\ifx\csname href\endcsname\relax
  \def\href#1#2{#2} \def\path#1{#1}\fi

\bibitem{Harland-Lang:2014zoa}
L.~A. Harland-Lang, A.~D. Martin, P.~Motylinski, R.~S. Thorne, {Parton
  distributions in the LHC era: MMHT 2014 PDFs}, Eur. Phys. J. C75~(5) (2015)
  204.
\newblock \href {http://arxiv.org/abs/1412.3989} {\path{arXiv:1412.3989}},
  \href {http://dx.doi.org/10.1140/epjc/s10052-015-3397-6}
  {\path{doi:10.1140/epjc/s10052-015-3397-6}}.

\bibitem{Abramowicz:2015mha}
H.~Abramowicz, et~al., {Combination of measurements of inclusive deep inelastic
  ${e^{\pm }p}$ scattering cross sections and QCD analysis of HERA data}, Eur.
  Phys. J. C75~(12) (2015) 580.
\newblock \href {http://arxiv.org/abs/1506.06042} {\path{arXiv:1506.06042}},
  \href {http://dx.doi.org/10.1140/epjc/s10052-015-3710-4}
  {\path{doi:10.1140/epjc/s10052-015-3710-4}}.

\bibitem{Alekhin:2017kpj}
S.~Alekhin, J.~Blümlein, S.~Moch, R.~Placakyte, {Parton distribution
  functions, $\alpha_s$, and heavy-quark masses for LHC Run II}, Phys. Rev.
  D96~(1) (2017) 014011.
\newblock \href {http://arxiv.org/abs/1701.05838} {\path{arXiv:1701.05838}},
  \href {http://dx.doi.org/10.1103/PhysRevD.96.014011}
  {\path{doi:10.1103/PhysRevD.96.014011}}.

\bibitem{Ball:2017nwa}
R.~D. Ball, et~al., {Parton distributions from high-precision collider data},
  Eur. Phys. J. C77~(10) (2017) 663.
\newblock \href {http://arxiv.org/abs/1706.00428} {\path{arXiv:1706.00428}},
  \href {http://dx.doi.org/10.1140/epjc/s10052-017-5199-5}
  {\path{doi:10.1140/epjc/s10052-017-5199-5}}.

\bibitem{Hou:2019qau}
T.-J. Hou, et~al., {Progress in the CTEQ-TEA NNLO global QCD analysis}\href
  {http://arxiv.org/abs/1908.11394} {\path{arXiv:1908.11394}}.

\bibitem{Altarelli:1978tq}
G.~Altarelli, G.~Martinelli, {Transverse Momentum of Jets in Electroproduction
  from Quantum Chromodynamics}, Phys.Lett. B76 (1978) 89.
\newblock \href {http://dx.doi.org/10.1016/0370-2693(78)90109-0}
  {\path{doi:10.1016/0370-2693(78)90109-0}}.

\bibitem{Kruczenski:2003be}
M.~Kruczenski, D.~Mateos, R.~C. Myers, D.~J. Winters, {Meson spectroscopy in
  AdS / CFT with flavor}, JHEP 0307 (2003) 049.
\newblock \href {http://arxiv.org/abs/hep-th/0304032}
  {\path{arXiv:hep-th/0304032}}.

\bibitem{Son:2003et}
D.~T.~Son, M.~A.~Stephanov, {QCD and dimensional deconstruction}, Phys.Rev. D69
  (2004) 065020.
\newblock \href {http://arxiv.org/abs/hep-ph/0304182}
  {\path{arXiv:hep-ph/0304182}}, \href
  {http://dx.doi.org/10.1103/PhysRevD.69.065020}
  {\path{doi:10.1103/PhysRevD.69.065020}}.

\bibitem{Kruczenski:2003uq}
M.~Kruczenski, D.~Mateos, R.~C. Myers, D.~J. Winters, {Towards a holographic
  dual of large N(c) QCD}, JHEP 0405 (2004) 041.
\newblock \href {http://arxiv.org/abs/hep-th/0311270}
  {\path{arXiv:hep-th/0311270}}, \href
  {http://dx.doi.org/10.1088/1126-6708/2004/05/041}
  {\path{doi:10.1088/1126-6708/2004/05/041}}.

\bibitem{Sakai:2004cn}
T.~Sakai, S.~Sugimoto, {Low energy hadron physics in holographic QCD},
  Prog.Theor.Phys. 113 (2005) 843--882.
\newblock \href {http://arxiv.org/abs/hep-th/0412141}
  {\path{arXiv:hep-th/0412141}}, \href {http://dx.doi.org/10.1143/PTP.113.843}
  {\path{doi:10.1143/PTP.113.843}}.

\bibitem{Erlich:2005qh}
J.~Erlich, E.~Katz, D.~T. Son, M.~A. Stephanov, {QCD and a holographic model of
  hadrons}, Phys.Rev.Lett. 95 (2005) 261602.
\newblock \href {http://arxiv.org/abs/hep-ph/0501128}
  {\path{arXiv:hep-ph/0501128}}, \href
  {http://dx.doi.org/10.1103/PhysRevLett.95.261602}
  {\path{doi:10.1103/PhysRevLett.95.261602}}.

\bibitem{Sakai:2005yt}
T.~Sakai, S.~Sugimoto, {More on a holographic dual of QCD}, Prog.Theor.Phys.
  114 (2005) 1083--1118.
\newblock \href {http://arxiv.org/abs/hep-th/0507073}
  {\path{arXiv:hep-th/0507073}}, \href {http://dx.doi.org/10.1143/PTP.114.1083}
  {\path{doi:10.1143/PTP.114.1083}}.

\bibitem{DaRold:2005zs}
L.~Da~Rold, A.~Pomarol, {Chiral symmetry breaking from five dimensional
  spaces}, Nucl.Phys. B721 (2005) 79--97.
\newblock \href {http://arxiv.org/abs/hep-ph/0501218}
  {\path{arXiv:hep-ph/0501218}}, \href
  {http://dx.doi.org/10.1016/j.nuclphysb.2005.05.009}
  {\path{doi:10.1016/j.nuclphysb.2005.05.009}}.

\bibitem{Brodsky:2014yha}
S.~J. Brodsky, G.~F. de~Teramond, H.~G. Dosch, J.~Erlich, {Light-Front
  Holographic QCD and Emerging Confinement}, Phys. Rept. 584 (2015) 1--105.
\newblock \href {http://arxiv.org/abs/1407.8131} {\path{arXiv:1407.8131}},
  \href {http://dx.doi.org/10.1016/j.physrep.2015.05.001}
  {\path{doi:10.1016/j.physrep.2015.05.001}}.

\bibitem{Maldacena:1997re}
J.~M. Maldacena, {The Large N limit of superconformal field theories and
  supergravity}, Adv.Theor.Math.Phys. 2 (1998) 231--252.
\newblock \href {http://arxiv.org/abs/hep-th/9711200}
  {\path{arXiv:hep-th/9711200}}.

\bibitem{Gubser:1998bc}
S.~Gubser, I.~R. Klebanov, A.~M. Polyakov, {Gauge theory correlators from
  noncritical string theory}, Phys.Lett. B428 (1998) 105--114.
\newblock \href {http://arxiv.org/abs/hep-th/9802109}
  {\path{arXiv:hep-th/9802109}}, \href
  {http://dx.doi.org/10.1016/S0370-2693(98)00377-3}
  {\path{doi:10.1016/S0370-2693(98)00377-3}}.

\bibitem{Witten:1998qj}
E.~Witten, {Anti-de Sitter space and holography}, Adv.Theor.Math.Phys. 2 (1998)
  253--291.
\newblock \href {http://arxiv.org/abs/hep-th/9802150}
  {\path{arXiv:hep-th/9802150}}.

\bibitem{Donnachie:1992ny}
A.~Donnachie, P.~Landshoff, {Total cross-sections}, Phys.Lett. B296 (1992)
  227--232.
\newblock \href {http://arxiv.org/abs/hep-ph/9209205}
  {\path{arXiv:hep-ph/9209205}}, \href
  {http://dx.doi.org/10.1016/0370-2693(92)90832-O}
  {\path{doi:10.1016/0370-2693(92)90832-O}}.

\bibitem{ForshawRoss}
J.~R. Forshaw, D.~A. Ross, Quantum Chromodynamics and the Pomeron, Cambridge
  University Press, Cambridge, 1997.

\bibitem{PomeronPhysicsandQCD}
S.~Donnachie, G.~Dosch, P.~Landshoff, O.~Nachtmann, Pomeron Physics and QCD,
  Cambridge University Press, Cambridge, 2002.

\bibitem{Polchinski:2001tt}
J.~Polchinski, M.~J. Strassler, {Hard scattering and gauge / string duality},
  Phys.Rev.Lett. 88 (2002) 031601.
\newblock \href {http://arxiv.org/abs/hep-th/0109174}
  {\path{arXiv:hep-th/0109174}}, \href
  {http://dx.doi.org/10.1103/PhysRevLett.88.031601}
  {\path{doi:10.1103/PhysRevLett.88.031601}}.

\bibitem{Polchinski:2002jw}
J.~Polchinski, M.~J. Strassler, {Deep inelastic scattering and gauge / string
  duality}, JHEP 0305 (2003) 012.
\newblock \href {http://arxiv.org/abs/hep-th/0209211}
  {\path{arXiv:hep-th/0209211}}.

\bibitem{BoschiFilho:2005yh}
H.~Boschi-Filho, N.~R.~F.~Braga, H.~L. Carrion, {Glueball Regge trajectories
  from gauge/string duality and the Pomeron}, Phys.Rev. D73 (2006) 047901.
\newblock \href {http://arxiv.org/abs/hep-th/0507063}
  {\path{arXiv:hep-th/0507063}}, \href
  {http://dx.doi.org/10.1103/PhysRevD.73.047901}
  {\path{doi:10.1103/PhysRevD.73.047901}}.

\bibitem{Brower:2006ea}
R.~C. Brower, J.~Polchinski, M.~J. Strassler, C.-I. Tan, {The Pomeron and
  gauge/string duality}, JHEP 0712 (2007) 005.
\newblock \href {http://arxiv.org/abs/hep-th/0603115}
  {\path{arXiv:hep-th/0603115}}, \href
  {http://dx.doi.org/10.1088/1126-6708/2007/12/005}
  {\path{doi:10.1088/1126-6708/2007/12/005}}.

\bibitem{Hatta:2007he}
Y.~Hatta, E.~Iancu, A.~Mueller, {Deep inelastic scattering at strong coupling
  from gauge/string duality: The Saturation line}, JHEP 0801 (2008) 026.
\newblock \href {http://arxiv.org/abs/0710.2148} {\path{arXiv:0710.2148}},
  \href {http://dx.doi.org/10.1088/1126-6708/2008/01/026}
  {\path{doi:10.1088/1126-6708/2008/01/026}}.

\bibitem{Brower:2007qh}
R.~C. Brower, M.~J. Strassler, C.-I. Tan, {On the eikonal approximation in AdS
  space}, JHEP 0903 (2009) 050.
\newblock \href {http://arxiv.org/abs/0707.2408} {\path{arXiv:0707.2408}},
  \href {http://dx.doi.org/10.1088/1126-6708/2009/03/050}
  {\path{doi:10.1088/1126-6708/2009/03/050}}.

\bibitem{BallonBayona:2007rs}
C.~Ballon~Bayona, H.~Boschi-Filho, N.~R. Braga, {Deep inelastic structure
  functions from supergravity at small x}, JHEP 0810 (2008) 088.
\newblock \href {http://arxiv.org/abs/0712.3530} {\path{arXiv:0712.3530}},
  \href {http://dx.doi.org/10.1088/1126-6708/2008/10/088}
  {\path{doi:10.1088/1126-6708/2008/10/088}}.

\bibitem{Brower:2007xg}
R.~C. Brower, M.~J. Strassler, C.-I. Tan, {On The Pomeron at Large 't Hooft
  Coupling}, JHEP 0903 (2009) 092.
\newblock \href {http://arxiv.org/abs/0710.4378} {\path{arXiv:0710.4378}},
  \href {http://dx.doi.org/10.1088/1126-6708/2009/03/092}
  {\path{doi:10.1088/1126-6708/2009/03/092}}.

\bibitem{Cornalba:2008sp}
L.~Cornalba, M.~S. Costa, {Saturation in Deep Inelastic Scattering from
  AdS/CFT}, Phys.Rev. D78 (2008) 096010.
\newblock \href {http://arxiv.org/abs/0804.1562} {\path{arXiv:0804.1562}},
  \href {http://dx.doi.org/10.1103/PhysRevD.78.096010}
  {\path{doi:10.1103/PhysRevD.78.096010}}.

\bibitem{Pire:2008zf}
B.~Pire, C.~Roiesnel, L.~Szymanowski, S.~Wallon, {On AdS/QCD correspondence and
  the partonic picture of deep inelastic scattering}, Phys.Lett. B670 (2008)
  84--90.
\newblock \href {http://arxiv.org/abs/0805.4346} {\path{arXiv:0805.4346}},
  \href {http://dx.doi.org/10.1016/j.physletb.2008.10.026}
  {\path{doi:10.1016/j.physletb.2008.10.026}}.

\bibitem{Cornalba:2010vk}
L.~Cornalba, M.~S. Costa, J.~Penedones, {AdS black disk model for small-x DIS},
  Phys.Rev.Lett. 105 (2010) 072003.
\newblock \href {http://arxiv.org/abs/1001.1157} {\path{arXiv:1001.1157}},
  \href {http://dx.doi.org/10.1103/PhysRevLett.105.072003}
  {\path{doi:10.1103/PhysRevLett.105.072003}}.

\bibitem{Brower:2010wf}
R.~C. Brower, M.~Djuric, I.~Sarcevic, C.-I. Tan, {String-Gauge Dual Description
  of Deep Inelastic Scattering at Small-$x$}, JHEP 1011 (2010) 051.
\newblock \href {http://arxiv.org/abs/1007.2259} {\path{arXiv:1007.2259}},
  \href {http://dx.doi.org/10.1007/JHEP11(2010)051}
  {\path{doi:10.1007/JHEP11(2010)051}}.

\bibitem{Watanabe:2012uc}
A.~Watanabe, K.~Suzuki, {Transition from soft- to hard-Pomeron in the structure
  functions of hadrons at small-$x$ from holography}, Phys.Rev. D86 (2012)
  035011.
\newblock \href {http://arxiv.org/abs/1206.0910} {\path{arXiv:1206.0910}},
  \href {http://dx.doi.org/10.1103/PhysRevD.86.035011}
  {\path{doi:10.1103/PhysRevD.86.035011}}.

\bibitem{Stoffers:2012zw}
A.~Stoffers, I.~Zahed, {Holographic Pomeron: Saturation and DIS}, Phys. Rev.
  D87 (2013) 075023.
\newblock \href {http://arxiv.org/abs/1205.3223} {\path{arXiv:1205.3223}},
  \href {http://dx.doi.org/10.1103/PhysRevD.87.075023}
  {\path{doi:10.1103/PhysRevD.87.075023}}.

\bibitem{Watanabe:2013spa}
A.~Watanabe, K.~Suzuki, {Nucleon structure functions at small $x$ via the
  Pomeron exchange in AdS space with a soft infrared wall}, Phys. Rev. D89~(11)
  (2014) 115015.
\newblock \href {http://arxiv.org/abs/1312.7114} {\path{arXiv:1312.7114}},
  \href {http://dx.doi.org/10.1103/PhysRevD.89.115015}
  {\path{doi:10.1103/PhysRevD.89.115015}}.

\bibitem{Agozzino:2013zgy}
L.~Agozzino, P.~Castorina, P.~Colangelo, {Nuclear Shadowing in the Holographic
  Framework}, Phys. Rev. Lett. 112~(4) (2014) 041601.
\newblock \href {http://arxiv.org/abs/1306.5072} {\path{arXiv:1306.5072}},
  \href {http://dx.doi.org/10.1103/PhysRevLett.112.041601}
  {\path{doi:10.1103/PhysRevLett.112.041601}}.

\bibitem{Watanabe:2015mia}
A.~Watanabe, H.-n. Li, {Photon structure functions at small $x$ in holographic
  QCD}, Phys. Lett. B751 (2015) 321--325.
\newblock \href {http://arxiv.org/abs/1502.03894} {\path{arXiv:1502.03894}},
  \href {http://dx.doi.org/10.1016/j.physletb.2015.10.069}
  {\path{doi:10.1016/j.physletb.2015.10.069}}.

\bibitem{Watanabe:2018owy}
A.~Watanabe, M.~Huang, {Total hadronic cross sections at high energies in
  holographic QCD}, Phys. Lett. B788 (2019) 256--260.
\newblock \href {http://arxiv.org/abs/1809.02515} {\path{arXiv:1809.02515}},
  \href {http://dx.doi.org/10.1016/j.physletb.2018.11.042}
  {\path{doi:10.1016/j.physletb.2018.11.042}}.

\bibitem{Xie:2019soz}
W.~Xie, A.~Watanabe, M.~Huang, {Elastic proton-proton scattering at LHC
  energies in holographic QCD}, JHEP 10 (2019) 053.
\newblock \href {http://arxiv.org/abs/1901.09564} {\path{arXiv:1901.09564}},
  \href {http://dx.doi.org/10.1007/JHEP10(2019)053}
  {\path{doi:10.1007/JHEP10(2019)053}}.

\bibitem{Abidin:2008hn}
Z.~Abidin, C.~E. Carlson, {Gravitational Form Factors in the Axial Sector from
  an AdS/QCD Model}, Phys.Rev. D77 (2008) 115021.
\newblock \href {http://arxiv.org/abs/0804.0214} {\path{arXiv:0804.0214}},
  \href {http://dx.doi.org/10.1103/PhysRevD.77.115021}
  {\path{doi:10.1103/PhysRevD.77.115021}}.

\bibitem{Abidin:2009hr}
Z.~Abidin, C.~E. Carlson, {Nucleon electromagnetic and gravitational form
  factors from holography}, Phys.Rev. D79 (2009) 115003.
\newblock \href {http://arxiv.org/abs/0903.4818} {\path{arXiv:0903.4818}},
  \href {http://dx.doi.org/10.1103/PhysRevD.79.115003}
  {\path{doi:10.1103/PhysRevD.79.115003}}.

\bibitem{CooperSarkar:1987ds}
A.~M. Cooper-Sarkar, G.~Ingelman, K.~Long, R.~Roberts, D.~Saxon, {MEASUREMENT
  OF THE LONGITUDINAL STRUCTURE FUNCTION AND THE SMALL x GLUON DENSITY OF THE
  PROTON}, Z.Phys. C39 (1988) 281.
\newblock \href {http://dx.doi.org/10.1007/BF01551005}
  {\path{doi:10.1007/BF01551005}}.

\bibitem{Barry:2018ort}
P.~C. Barry, N.~Sato, W.~Melnitchouk, C.-R. Ji, {First Monte Carlo Global QCD
  Analysis of Pion Parton Distributions}, Phys. Rev. Lett. 121~(15) (2018)
  152001.
\newblock \href {http://arxiv.org/abs/1804.01965} {\path{arXiv:1804.01965}},
  \href {http://dx.doi.org/10.1103/PhysRevLett.121.152001}
  {\path{doi:10.1103/PhysRevLett.121.152001}}.

\bibitem{Collaboration:2010ry}
F.~D. Aaron, et~al., {Measurement of the Inclusive $e^{\pm}p$ Scattering Cross
  Section at High Inelasticity $y$ and of the Structure Function $F_L$}, Eur.
  Phys. J. C71 (2011) 1579.
\newblock \href {http://arxiv.org/abs/1012.4355} {\path{arXiv:1012.4355}},
  \href {http://dx.doi.org/10.1140/epjc/s10052-011-1579-4}
  {\path{doi:10.1140/epjc/s10052-011-1579-4}}.

\bibitem{Henningson:1998cd}
M.~Henningson, K.~Sfetsos, {Spinors and the AdS / CFT correspondence},
  Phys.Lett. B431 (1998) 63--68.
\newblock \href {http://arxiv.org/abs/hep-th/9803251}
  {\path{arXiv:hep-th/9803251}}, \href
  {http://dx.doi.org/10.1016/S0370-2693(98)00559-0}
  {\path{doi:10.1016/S0370-2693(98)00559-0}}.

\bibitem{Muck:1998iz}
W.~Muck, K.~Viswanathan, {Conformal field theory correlators from classical
  field theory on anti-de Sitter space. 2. Vector and spinor fields}, Phys.Rev.
  D58 (1998) 106006.
\newblock \href {http://arxiv.org/abs/hep-th/9805145}
  {\path{arXiv:hep-th/9805145}}, \href
  {http://dx.doi.org/10.1103/PhysRevD.58.106006}
  {\path{doi:10.1103/PhysRevD.58.106006}}.

\bibitem{Contino:2004vy}
R.~Contino, A.~Pomarol, {Holography for fermions}, JHEP 0411 (2004) 058.
\newblock \href {http://arxiv.org/abs/hep-th/0406257}
  {\path{arXiv:hep-th/0406257}}, \href
  {http://dx.doi.org/10.1088/1126-6708/2004/11/058}
  {\path{doi:10.1088/1126-6708/2004/11/058}}.

\bibitem{Hong:2006ta}
D.~K. Hong, T.~Inami, H.-U. Yee, {Baryons in AdS/QCD}, Phys.Lett. B646 (2007)
  165--171.
\newblock \href {http://arxiv.org/abs/hep-ph/0609270}
  {\path{arXiv:hep-ph/0609270}}, \href
  {http://dx.doi.org/10.1016/j.physletb.2007.01.030}
  {\path{doi:10.1016/j.physletb.2007.01.030}}.

\bibitem{Aaron:2009aa}
F.~Aaron, et~al., {Combined Measurement and QCD Analysis of the Inclusive
  $e^{\pm}p$ Scattering Cross Sections at HERA}, JHEP 1001 (2010) 109.
\newblock \href {http://arxiv.org/abs/0911.0884} {\path{arXiv:0911.0884}},
  \href {http://dx.doi.org/10.1007/JHEP01(2010)109}
  {\path{doi:10.1007/JHEP01(2010)109}}.

\bibitem{James:1975dr}
F.~James, M.~Roos, {Minuit: A System for Function Minimization and Analysis of
  the Parameter Errors and Correlations}, Comput. Phys. Commun. 10 (1975)
  343--367.
\newblock \href {http://dx.doi.org/10.1016/0010-4655(75)90039-9}
  {\path{doi:10.1016/0010-4655(75)90039-9}}.

\bibitem{Furmanski:1981cw}
W.~Furmanski, R.~Petronzio, {Lepton - Hadron Processes Beyond Leading Order in
  Quantum Chromodynamics}, Z. Phys. C11 (1982) 293.
\newblock \href {http://dx.doi.org/10.1007/BF01578280}
  {\path{doi:10.1007/BF01578280}}.

\bibitem{Gluck:1999xe}
M.~Gluck, E.~Reya, I.~Schienbein, {Pionic parton distributions revisited},
  Eur.Phys.J. C10 (1999) 313--317.
\newblock \href {http://arxiv.org/abs/hep-ph/9903288}
  {\path{arXiv:hep-ph/9903288}}, \href
  {http://dx.doi.org/10.1007/s100529900124} {\path{doi:10.1007/s100529900124}}.

\end{thebibliography}

\end{document}